\documentclass[prb,aps,twocolumn,showkeys,showpacs,amsfonts,amsmath,amssymb,showkeys,floatfix,sort&compress,a4paper,byrevtex]{revtex4-1}

\usepackage{graphicx}
\usepackage{url}
\usepackage{ulem}

\usepackage{color}

\begin{document}

\title{Localization of metallicity and magnetic properties of graphene and of graphene nanoribbons doped with boron clusters}

\author{Cem \"{O}zdo\u{g}an$^{1,*}$, Jens Kunstmann$^{2,3}$, Alexander Quandt$^{4}$}
\affiliation{$^{1}$Department of Materials Science and Engineering, \c{C}ankaya University, 06530 Ankara, Turkey}
\email[Corresponding Author, Electronic address: ]{ozdogan@cankaya.edu.tr}
\affiliation{$^{2}$Institute for Materials Science and Max Bergmann Center of Biomaterials, TU Dresden, 01062 Dresden, Germany}
\affiliation{$^{3}$Department of Chemistry, Columbia University, 3000 Broadway New York, NY 10027, USA}
\affiliation{$^{4}$School of Physics and DST/NRF Centre of Excellence In Strong Materials, University of the Witwatersrand, Wits 2050, South Africa}

\date{\today}

\begin{abstract}
As a possible way of modifying the intrinsic properties of graphene we study the doping of graphene by embedded  boron clusters with density functional theory. Cluster doping is technologically relevant as the cluster implantation technique can be readily applied to graphene.
We find that B$_7$ clusters embedded into graphene and graphene nanoribbons are structurally stable and locally metallize the system. This is done both by the reduction of the Fermi energy and by the introduction of boron states near the Fermi level. A linear chain of boron clusters forms a metallic ``wire'' inside the graphene matrix. In a zigzag edge graphene nanoribbon the cluster-related states tend to  hybridize with the edge and bulk states.
The magnetism in boron doped graphene systems is generally very weak. The presence of boron clusters weakens the edge magnetism in zigzag edge graphene nanoribbon, rather than making the system appropriate for spintronics.
Thus the doping of graphene with the cluster implantation technique might be a viable technique to locally metallize graphene without destroying its attractive bulk properties.
\end{abstract}



\maketitle

\section{Introduction}

The two-dimensional (2D) honeycomb lattice of graphene is the basic building block of 0D fullerenes, 1D nanotubes or 3D graphite. Besides being a 2D topological structure with high strength and flexibility, graphene exhibits fascinating, and rather unique electronic properties, such as behaving like massless Dirac fermions and showing a particular high charge carrier mobility and electrical conductivity \cite{Novoselov2005b,Zhang2005,Katsnelson2006,Geim2007a,Novoselov2007a}. These exceptional properties lead to huge scientific interest in graphene, and they make this particular material a promising candidate for future technological applications \cite{CastroNeto2009,Singh2011,Mas-Balleste2011,Molitor2011}.

A free-standing graphene sheet is a semimetal. However many applications require semi-conducting properties. A band gap can be induced in graphene systems, for example by oxidation or by considering graphene nanoribbons (GNRs). Narrow width ribbons ($< ~10~ nm$) have a band gap due to lateral quantum confinement and edge effects \cite{Ezawa2006,Han2007,Li2008,Ritter2009}. A process for fabricating dense graphene nanoribbon arrays with widths as narrow as 10 nm has been developed quite recently \cite{Liu2012}, using self-assembled patterns of block copolymers on graphene, which was grown epitaxially on SiC.

The edge geometries of GNRs can be rather diverse \cite{Wassmann2008}. Usually armchair or zigzag types (AGNR or ZGNR) are considered in the literature \cite{Kobayashi2005,Jian2006,Kobayashi2006,Niimi2006,Girit2009,Liu2009,Reddy2009}. AGNRs are either metallic or semi-conducting depending on the ribbon width, and ZGNRs have metallic properties with flat energy bands near the Fermi level $E_F$. The latter give rise to a sharp peak in the electronic density of states (DOS) right at $E_F$. It is currently believed that this peak at $E_F$ is avoided by magnetic interactions, and that the ground state of ZGNRs is actually an antiferromagnetic semi-conductor (AFM, atoms are ferromagnetically ordered along one edge and antiferromagnetically ordered between opposite edges) with a small band gap. In density functional theory (DFT) calculations the AFM magnetic ground state is almost degenerate to a ferromagnetic (FM) state \cite{Fujita1996,Nakada1996}. These results suggested the use of ZGNRs in spintronic 
applications, and they inspired tremendous theoretical work.
It is understood now that there should either be some translational symmetry breaking between edges in GNRs or some local sublattice imbalances in graphene/graphite to create localized states at $E_F$ to obtain a finite magnetic moment. Several approaches have been proposed, such as the introduction of vacancies \cite{Lehtinen2004,Yazyev2007,Palacios2008,Esquinazi2010,Faccio2010,Ugeda2010}, networks of point defects \cite{Cervenka2009}, impurities/chemical doping \cite{Martins2007,Lherbier2008,Quandt2008a,Quandt2008b,Panchokarla2009,Pontes2009,Wei2009,Dai2010,Faccio2010,Yu2010,Wang2010,Padilha2011,Power2011,Guodong2012,Dai2010a,Wang2010a,Wang2009,Wang2012a,Uchoa2008a,Miwa2010,Huang2007}, partial hydrogenation \cite{Soriano2011,Xie2011}, edge roughness \cite{Wimmer2008}, electron or hole doping \cite{Jung2009b} or an applied electric field to introduce half-metallicity \cite{Son2006a}.
However, some recent studies also point out that the edge states and the associated edge magnetism might be either very weak, or absent, or even artefacts of improper theoretical methodology \cite{Kunstmann2011,Deleuze2012,Martinez-Martin2010}.

Another way of modulating and controlling the electronic properties of graphene is chemical doping, which was widely studied in the past \cite{Martins2007,Lherbier2008,Quandt2008a,Quandt2008b,Panchokarla2009,Pontes2009,Wei2009,Faccio2010,Yu2010,Wang2010,Padilha2011,Dai2010a,Wang2010a,Wang2009,Wang2012a,Uchoa2008a,Miwa2010,Huang2007}. The synthesis of substitutionally doped graphene samples is presently reported in the literature, such as B- and N-doping by arc discharge \cite{Panchokarla2009}, N doping by chemical vapor deposition (CVD) \cite{Wei2009}  methods, and doping with transition metals atoms \cite{Guodong2012}. B/N- \cite{Panchokarla2009,Wang2010a}, B- \cite{Pontes2009,Faccio2010,Martins2007}, P-\cite{Dai2010}, transition metals atoms \cite{Uchoa2008a} doped graphene systems were also investigated from a theoretical point of view, with particular focus on the effects of substitutional doping on the structural and electronic properties of graphene. Boron, the neighbor of carbon in the periodic table,
 seems to be an ideal dopant \cite{Martins2007}. However, individual boron atoms in graphene are rather mobile and tend to aggregate at the edges and not in the bulk of the sample, as studies of boron doped graphene nanoribbons \cite{Martins2007} and boron doped carbon nanotubes \cite{Hernandez2000} show. In the present article we do not suggest to dope graphene with individual boron atoms, but with entire boron clusters.

Small boron clusters have a quasiplanar geometry and can be viewed as an assembly of pyramidal $B_7$ building blocks (Aufbau principle) \cite{Boustani1997}. Boron and carbon have roughly the same covalent radius. Therefore, one may expect that boron could easily be integrated into carbon systems. On the other hand, nitrogen adsorption causes some structural distortions on graphene, despite having a similar atomic radius to carbon as well \cite{Pontes2009}.

We have previously reported that boron cluster can be easily accepted by graphene \cite{Quandt2008b}.
Ion beam methods can be utilized efficiently to dope boron into the graphene/graphite matrix, either as atomic sites \cite{Bangert2010,Ahlgren2011} or as clusters \cite{Nakashima2011}. A cluster implanter particularly suited for boron ($B_{10}$, $B_{18}$ clusters) has been developed with the enhancements of ultra low energy beam currents \cite{Nakashima2011}. Studies of doping of few-layered graphene and carbon nanotubes for a range of dopants (including boron) using ion implantation techniques \cite{Bangert2010} , as well as atomistic simulation of low energy boron and nitrogen ions implantation into graphene \cite{Ahlgren2011} have recently been reported in the literature.

In the present work we study the effect of embedded boron clusters on the electronic and magnetic properties of graphene systems. We consider 2D graphene and 1D AGNRs and ZGNRs. These systems are substitutionally doped with B$_7$ clusters to create (1) localized states, (2) local sublattice imbalances and (3) localized magnetic moments. We find that the B$_7$ clusters locally metallize the graphene matrix, and that linear chains of boron clusters form metallic wires inside the graphene matrix. Furthermore the presence of the B$_7$ clusters in magnetic ZGNR weakens the magnetism in such systems.

\section{Computational Method}
\label{sec:Computational Method}

Our calculations are based on the density functional theory \cite{Kohn1965} (DFT) within the framework of the generalized gradient approximation (GGA) using the Perdew-Wang parametrization (PW91) \cite{Perdew1992,Perdew1992a}. We used the Vienna ab initio simulation package (VASP, version 4.6) \cite{Kresse1996,Kresse1996a} employing the projected augmented wave (PAW) method \cite{Blochl1994,Kresse1999}.

In our study, we consider graphene supercells, 9-AGNR(+H) and 10-ZGNR(+H)s and their boron doped counterparts (+BD, +isolated-BD). The prefix ``9 (10)'' represents the number of armchair (zigzag) lines cross the width of the ribbon, and the suffix ``+H'' indicates that  each carbon edge atom is passivated by one hydrogen atom. A suffix ``+BD'' that the structure is boron doped and ``+isolated-BD'' indicates that the embedded boron clusters in adjacent unit cells of the graphene supercell systems are separated by at least 10 \AA.

A supercell system still repeats itself periodically in both the horizontal (x-) and the vertical (y-) direction. In order to simulate GNRs, and in order to avoid any residual interactions, each replica is separated from its neighboring replica by least 10 \AA~in both, the edge--edge (to the right or to the left) and the (z-) direction perpendicular to the planes of the replicas
\footnote
{
The optimized lattice parameters of the corresponding unit cells in their (magnetic) ground states are given as follows:
graphene: A=2.460, B=2.459, C=10 \AA,
graphene supercell: A=21.972, B=8.458, C=10 \AA,
graphene+isolated-BD supercell: A=22.140, B=17.043, C=10 \AA,
graphene+BD supercell: A=22.140, B=8.522, C=10 \AA,
For all AGNRs A=43$\dots$47 \AA, C=10 \AA,
9-AGNR, 9-AGNR+H: B=8.509 \AA,
9-AGNR+BD, 9-AGNR+BD+H: B=8.522 \AA,
For all ZGNRs B=35$\dots$37 \AA, C=10 \AA,
10-ZGNR: A=7.416 \AA,
10-ZGNR+H: A=7.383 \AA,
10-ZGNR+BD, 10-ZGNR+BD+H: A=7.511 \AA,
}.

A plane wave basis set with a kinetic energy cutoff value of 450 eV is used for all calculations. For the k-point sampling of the Brillouin zone, a $\Gamma$-point centered Monkhorst-Pack grid was used. The k-space integration was carried out with the method of Methfessel and Paxton in first order \cite{Methfessel1989} and a smearing width of 0.1 eV. Optimal k-point meshes are individually converged for each system by reducing the changes in the total energy at least below 10 meV/atom. \footnote
{
Optimized k-point meshes for the different systems at their (magnetic) ground states are;
graphene: 10x10x3,
graphene supercell: 3x3x3,
graphene+isolated-BD supercell: 3x3x3,
graphene+BD supercell: 3x5x2,
9-AGNR: 2x5x3,
9-AGNR+H: 3x3x3,
9-AGNR+BD: 2x5x3,
9-AGNR+BD+H: 3x3x3,
10-ZGNR: 4x3x3,
10-ZGNR+H: 3x2x3,
10-ZGNR+BD: 3x3x3,
10-ZGNR+BD+H: 5x2x3.
}
After obtaining the optimal k-point meshes, the geometries of all systems were fully optimized (relaxed) in a self-consistent fashion. The relaxed structures were close to the unrelaxed ideal geometries. After the relaxation of each structure an additional static self-consistent calculation was performed to obtain the charge density and the total energies of the system, where $k$-space integrations were carried out using the tetrahedron method \cite{Blochl1994a}. In all of our self-consistent calculations the total energies were converged, such that energetic changes were less than 10 meV. Band structures and DOS of the optimized systems were computed non-self-consistently by using the charge densities obtained in the final self-consistent calculation. At least twice the optimal size of $k$-point meshes was used for the DOS calculations. For magnetic systems collinear, spin polarized (SP) DFT calculations were performed in all the steps described above. The different magnetic states were determined by 
providing different initial spin densities at the beginning of the self-consistency cycles. System sizes, total energies, and magnetic moments of the systems considered for this study are summarized in table \ref{tab:mag_energies}. In the following, the non-spin-polarized calculations will be abbreviated by NSP.

\section{Results and Discussion}
In the presented study, a seven atom planar boron B$_7$ cluster is embedded into graphene and various GNRs (see for example Fig.~\ref{fig:10-zgnr_borondoped+HT_PCD_TotalDOS_PlainBands_SpinDensities}(h)). The basic stability of this structure was reported by us in an earlier study \cite{Quandt2008a,Quandt2008b}.
The boron atom, being one electron short in comparison to the carbon atom, is a p-type dopant. Embedding a $B_7$ unit (21 valence electrons) into a graphene system leads to the replacement of a $C_6$ hexagon (24 valence electrons), which results in a decrease of 3 electrons (introduction of 3 holes) per unit cell, and thus in the lowering of the Fermi energy $E_F$.
The degree of doping is calculated with the following equation
\begin{eqnarray}
\Delta E_\mathrm{F} = E_\mathrm{F}^\mathrm{pristine} - E_\mathrm{F}^\mathrm{doped},
\label{eqn:E_Fermi}
\end{eqnarray}
and numerical values for the shift of the Fermi energy are tabulated in Table \ref{tab:mag_energies}.

The atomic structures together with partial charge densities (PCD), projected DOSs, plain band structures and spin densities of an optimized graphene supercell and of GNR+H systems with boron cluster doping are shown in Figs.~\ref{fig:graphene_borondoped-isolated_PCD_TotalDOS_PlainBands_SpinDensities}-\ref{fig:10-zgnr_borondoped+HT_PCD_TotalDOS_PlainBands_SpinDensities}(a)-(d), respectively. 
The PCD is defined as
\begin{eqnarray}
\mathrm{PCD}(\mathbf{r}) = \sum_{n \mathbf{k}} \int_{E_\mathrm{F} - \epsilon}^{E_\mathrm{F} + \epsilon} dE \ \delta(E - E_{n \mathbf{k}}) |\phi_{n \mathbf{k}}(\mathbf{r})|^2,
\label{eqn:PCD}
\end{eqnarray}
where $\phi_{n \mathbf{k}}(\mathbf{r})$ are the Kohn-Sham orbitals with the band index $n$ and the wave vector $\mathbf{k}$, $E_{n \mathbf{k}}$ are the band energies, the summation over $\mathbf{k}$ represents a Brillouine zone average, and the parameter $\epsilon = 0.1$ eV in all our calculations. The PCD visualizes the localization of electronic states and it is drawn on top of the atomic structure. Note that the chosen energy range is relevant for electric conduction.

\subsection{Localization of Metallicity}
\label{sec:Localization of metallicity}

\subsubsection{Isolated Boron Cluster in Graphene}
\label{sec:Isolated Boron cluster in Graphene}

\begin{figure}[h!]
\includegraphics[width=.96\columnwidth,angle=0]{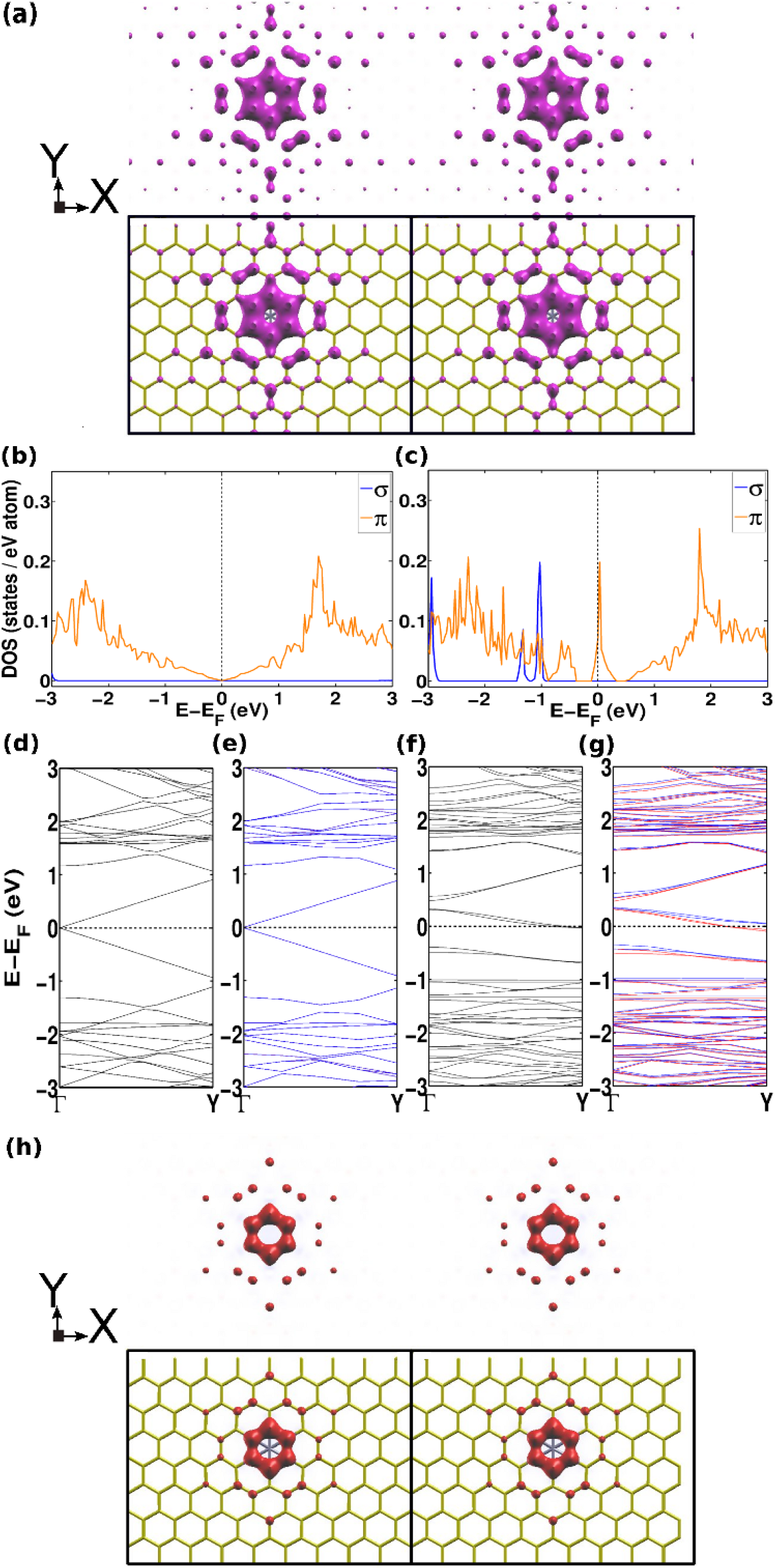}
\caption{\label{fig:graphene_borondoped-isolated_PCD_TotalDOS_PlainBands_SpinDensities}
(Color online) Electronic structure of graphene with embedded B$_7$ clusters that are (nearly) isolated from each other (graphene+isolated-BD); the center-to-center separation of the boron clusters in the y-direction is 17 \AA. (a) Non-spin-polarized (NSP) partial charge density (PCD) for states in the energy window $|E-E_F| < 0.1$ eV at contours of 0.01 e/\AA$^3$; the PCD is basically localized on the outer boron atoms of the cluster as well as on nearby carbon atoms. (b,c) NSP density of states (DOS) of pristine and boron doped systems, respectively; orange (light gray) represents the $\pi$ character and blue (dark gray) represents the $\sigma$ character. (d-g) Band structures for NSP pristine, spin-polarized (SP) pristine, NSP boron doped and SP boron doped cases, respectively. (h) The spin density at contours of 0.01 $\mu_\mathrm{B}$/\AA$^3$ is localized on the cluster. In (a) and (h) four unit cells are depicted.
}
\end{figure}

First we study the properties of an nearly isolated  B$_7$ cluster in graphene (graphene+isolated-BD system). We use a relatively large supercell to ensure that the periodic replica of the clusters do not interact\footnote
{
The center-to-center separation of the clusters for the different systems in x-,y-directions are;
graphene+isolated-BD supercell: 22.14 \AA, 17.04 \AA,
graphene+BD supercell: $\sim$22.14 \AA, $\sim$8.52 \AA,
9-AGNR+BD+H: 8.52 \AA~ in y-direction,
10-ZGNR+BD+H: 7.38 \AA~ in x-direction.
}.
The optimized structure together with the PCD in the specified energy window is depicted in Figure \ref{fig:graphene_borondoped-isolated_PCD_TotalDOS_PlainBands_SpinDensities}(a).
The PCD is basically localized on the outer boron atoms of the cluster and on nearby carbon atoms, whereas the central atom does not contribute to the states near $E_F$. These electronic states have $\pi$ character, as seen from the projected DOS plot in Figure \ref{fig:graphene_borondoped-isolated_PCD_TotalDOS_PlainBands_SpinDensities}(c). The $\sigma$ states are responsible for in-plane stability of boron clusters within a graphene matrix, and they do not contribute to the states near $E_F$.
The general form of the DOS of the pristine and the boron doped systems in Figs. \ref{fig:graphene_borondoped-isolated_PCD_TotalDOS_PlainBands_SpinDensities}(b,c) is quite similar, indicating that the boron clusters do not significantly alter the bulk properties of the system in this energy window. Essential changes happen near $E_F$, where the DOS has a pronounced peak. The peaks indicates that the induced boron states are localize,  which is also confirmed by looking at the PCD.

The change of the band structures from Fig.~\ref{fig:graphene_borondoped-isolated_PCD_TotalDOS_PlainBands_SpinDensities}(d) to (f) can be described as twofold: (i) creation of new states near $E_F$ that are primarily localized on the boron atoms (as is obvious from the PCD plot shown above) and (ii) reduction of $E_F$ by $\Delta E_\mathrm{F}$ ca.~0.2 eV (see Tab.~\ref{tab:mag_energies}) in a way, that cannot be described by the rigid band model.
Though the system has a finite DOS($E_F$), and despite the fact that we see bands crossing $E_F$ in Fig.~\ref{fig:graphene_borondoped-isolated_PCD_TotalDOS_PlainBands_SpinDensities}(f), the system is not necessarily a metal, because the presumed conduction channels near $E_F$ are highly localized and have little overlap. However we can conclude that the boron cluster \textit{locally} metallizes the system. In order to create pronounced metallic conduction channels one has to decrease the distance between the clusters. This is what we will do in the next subsection.

\subsubsection{Boron Cluster Chain in Graphene}
\label{sec:Boron cluster chain in Graphene}

\begin{figure}[h!]
\includegraphics[width=\columnwidth,angle=0]{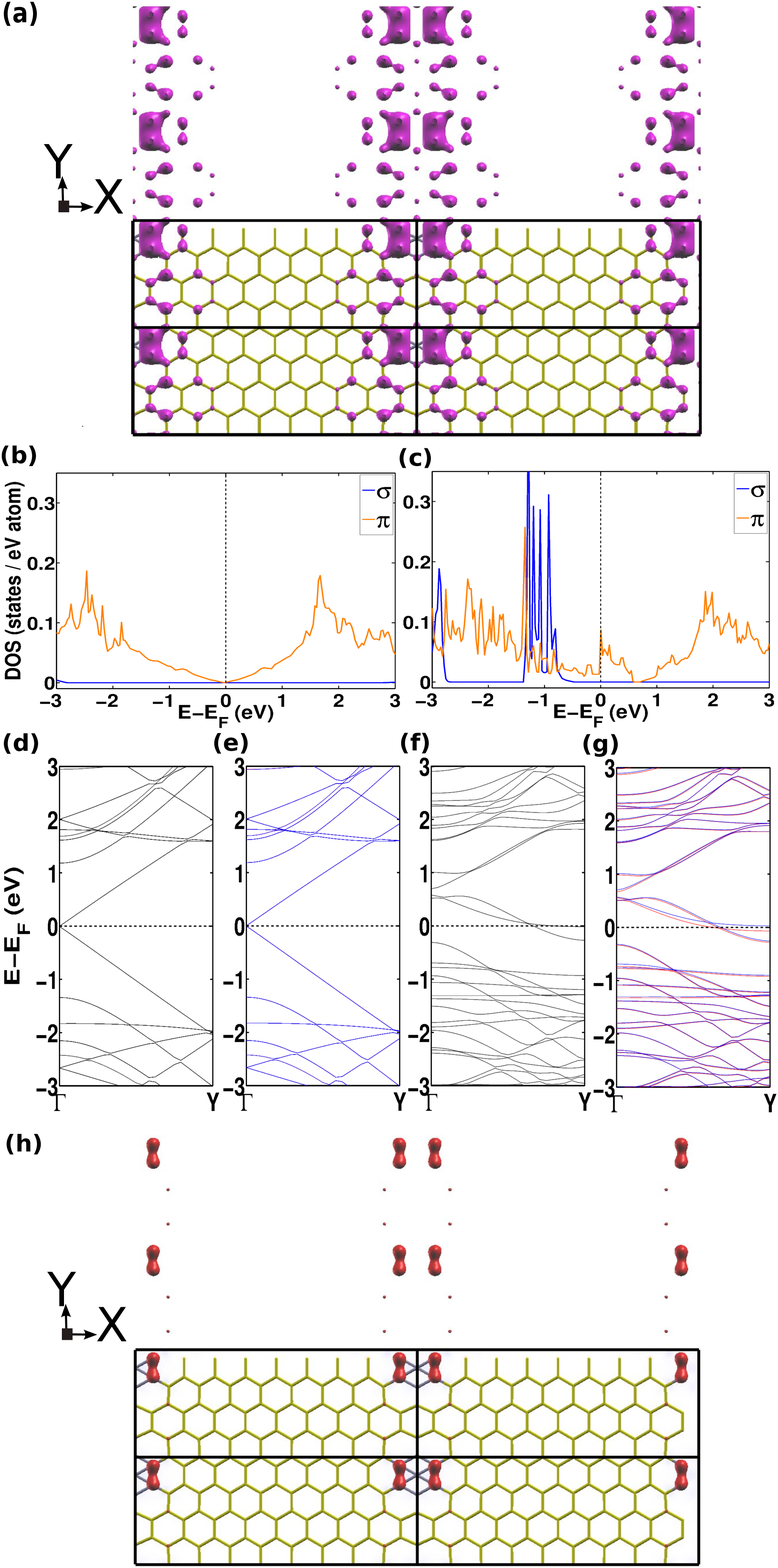}
\caption{\label{fig:graphene_borondoped_PCD_TotalDOS_PlainBands_SpinDensities}
(Color online) Electronic structure of graphene with B$_7$ clusters forming a linear chain with a center-to-center separation of 8.5 \AA (graphene+BD).
(a) The NSP PCD indicates that the states near $E_F$ form a 1D metallic ``wire'' inside the 2D graphene matrix. (b,c) NSP DOSs of pristine and boron doped cases, respectively.  (d-g) plain band structures of NSP pristine, SP pristine, NSP boron doped and SP boron doped cases, respectively. (h) The spin density shows that the magnetic moments on the different clusters are smaller than the ones on the isolated clusters in Fig.~\ref{fig:graphene_borondoped-isolated_PCD_TotalDOS_PlainBands_SpinDensities}.
Eight unit cells are depicted in (a) and (h). Color codes, energy ranges, contours and abbreviations are the same as in Fig.~\ref{fig:graphene_borondoped-isolated_PCD_TotalDOS_PlainBands_SpinDensities}.
}
\end{figure}

In the graphene+BD system the distance between the clusters along the y-direction is halved in comparison to the previous case. The atomic structure along y can be described as an alternating B$_7$-C$_6$ chain. The geometry and the corresponding PCD are depicted in Figure \ref{fig:graphene_borondoped_PCD_TotalDOS_PlainBands_SpinDensities}(a).
The  PCD indicates that the states near $E_F$ form a 1D metallic ``wire'' inside the 2D graphene matrix. The overlap between states related to the cluster tends to increase the dispersion of the the bands near $E_F$ in Fig.~\ref{fig:graphene_borondoped_PCD_TotalDOS_PlainBands_SpinDensities}(f)). This overlap also reduces the peak in the DOS($E_F$) in Fig.~\ref{fig:graphene_borondoped_PCD_TotalDOS_PlainBands_SpinDensities}(c). As the cluster density is doubled compared to the previously considered case the doping level is also approximately doubled to $\Delta E_\mathrm{F}$ ca.~0.4 eV.
The presence of boron $\sigma$ states in the occupied part of the DOS (at -1 and -3 eV) is qualitatively similar to the previous case, but this time it is much more pronounced.

\subsubsection{Boron cluster chain in AGNR}
\label{sec:Boron cluster chain in AGNR}
\begin{figure}[h!]
\includegraphics[width=\columnwidth,angle=0]{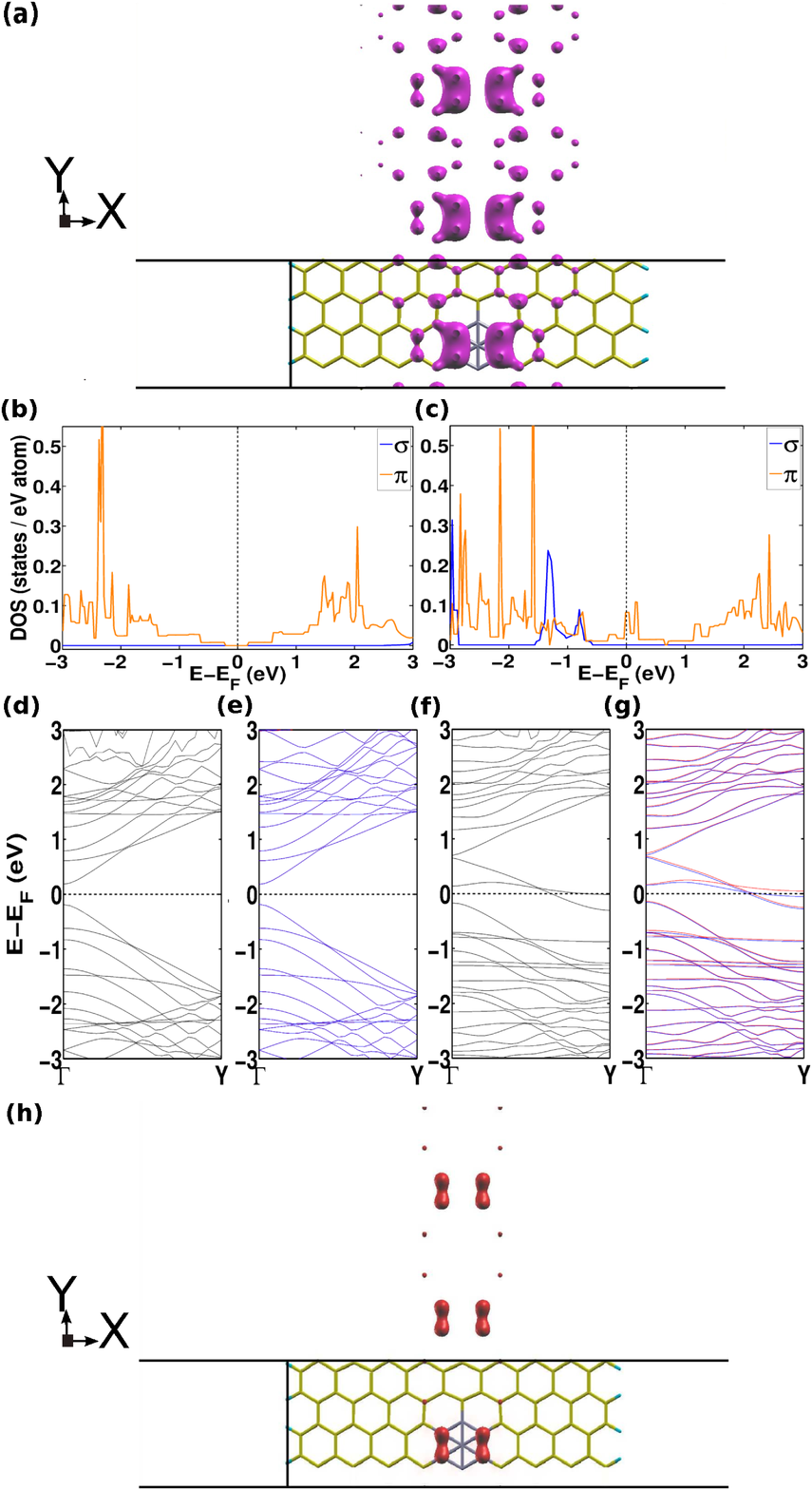}
\caption{\label{fig:9-agnr_borondoped+HT_PCD_TotalDOS_PlainBands_SpinDensities}
(Color online) Electronic structure of B$_7$ clusters forming a linear chain within an armchair graphene nanoribbon that has hydrogen terminated edges (9-AGNR+BD+H)
(a) The NSP PCD is almost identical to the one in Fig.~\ref{fig:graphene_borondoped_PCD_TotalDOS_PlainBands_SpinDensities}(a), i.e., the states form a  metallic ``wire'' inside the nanoribbon. (b,c) NSP  DOSs of pristine and boron doped cases. (d-g) Plain band structures for NSP pristine, SP pristine, NSP boron doped and SP boron doped cases, respectively. (h) spin density at contours of iso = -0.01 $\mu_\mathrm{B}$/\AA$^3$. Three unit cells are depicted in (a) and (h). Color codes, energy ranges, contours and abbreviations are the same as in Fig.~\ref{fig:graphene_borondoped-isolated_PCD_TotalDOS_PlainBands_SpinDensities}.
}
\end{figure}

The PCD of the alternating B$_7$-C$_6$ chain in an 9-AGNR+H is shown for the optimized geometry  in Figure \ref{fig:9-agnr_borondoped+HT_PCD_TotalDOS_PlainBands_SpinDensities}(a).
A comparison with  Figure \ref{fig:graphene_borondoped_PCD_TotalDOS_PlainBands_SpinDensities}(a) indicates that the PCDs of the two systems are almost identical. Thus the finite size of the AGNR does not influence the basic electronic properties of the system close to $E_F$. This further supports our finding that the metallization of the considered graphene systems by the boron clusters tends to happens on a very small length scale. The undoped 9-AGNR+H system is a semi-conductor with a band gap of 0.37 eV (see Fig.~\ref{fig:9-agnr_borondoped+HT_PCD_TotalDOS_PlainBands_SpinDensities}(d)). Doping and appearance of boron related $\pi$ and $\sigma$ states turn the system into a 1D metal with 4 (3) bands at $E_F$ for the NSP (SP) cases (this band counting includes the spin degeneracy).
The $\sigma$ states at about -1 eV in Fig.~\ref{fig:9-agnr_borondoped+HT_PCD_TotalDOS_PlainBands_SpinDensities}(c) are now much more smeared out in energy compared to the previous systems.

\subsubsection{Boron cluster chain in ZGNR}
\label{sec:Boron cluster chain in ZGNR}

\begin{figure}[h!]
\includegraphics[width=\columnwidth,angle=0]{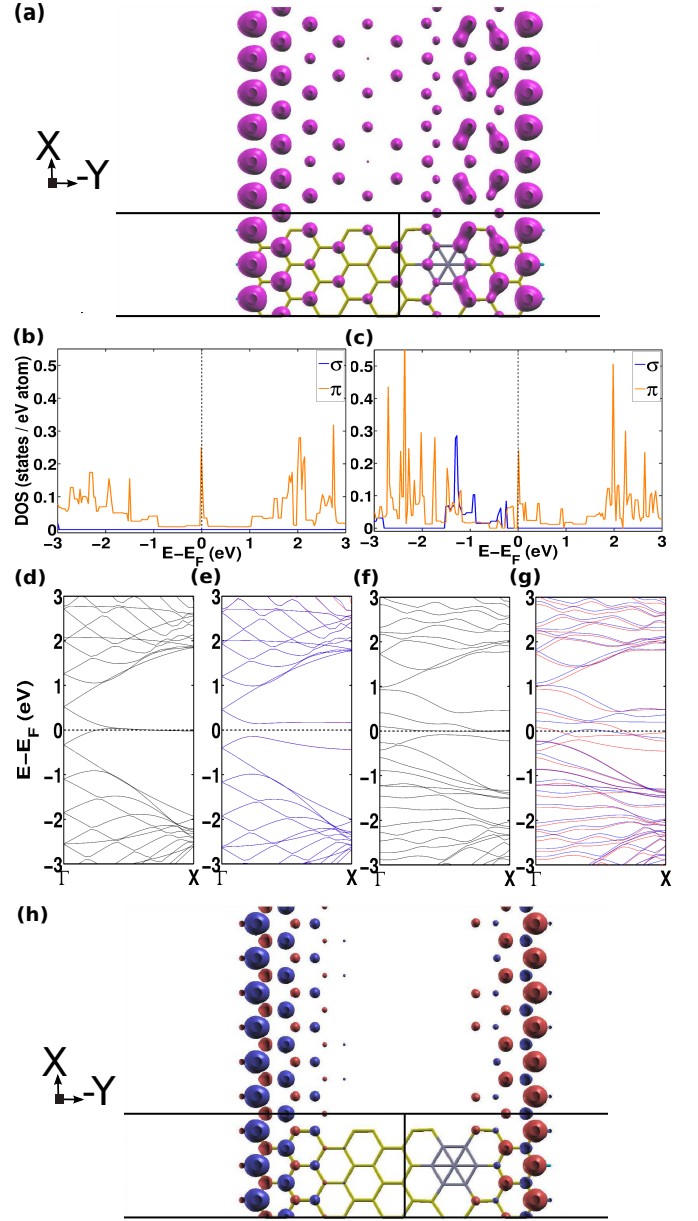}
\caption{\label{fig:10-zgnr_borondoped+HT_PCD_TotalDOS_PlainBands_SpinDensities}
(Color online) Electronic structure of B$_7$ clusters forming a linear chain within a zigzag graphene nanoribbon that has  hydrogen terminated edges (10-ZGNR+BD+H) (a) The NSP PCD shows that the states from the boron cluster hybridize with the edge and bulk states. (b,c) NSP  DOSs of pristine and boron doped cases. (d-g) Band structures for NSP pristine, antiferromagnetically (AFM) SP pristine, NSP boron doped and ferromagnetically (FM) SP  boron doped cases, respectively. (h) The spin density of the AFM state shows that magnetic moments on and in the vicinity of the cluster are suppressed. Thus the presence of the B$_7$ clusters weakens the edge magnetism. Three unit cells are depicted in (a) and (h). Color codes, energy ranges, contours and abbreviations are the same as in Fig.~\ref{fig:graphene_borondoped-isolated_PCD_TotalDOS_PlainBands_SpinDensities}.
}
\end{figure}
The undoped 10-ZGNR+H exhibits a considerable peak in the NSP DOS($E_F$) (see Fig.~\ref{fig:10-zgnr_borondoped+HT_PCD_TotalDOS_PlainBands_SpinDensities}(b)) coming from the flat bands at $E_F$ in the NSP band structure (see Fig.~\ref{fig:10-zgnr_borondoped+HT_PCD_TotalDOS_PlainBands_SpinDensities}(d)).
This peak is associated with the edge states mentioned in the Introduction.
As discernible in Figure \ref{fig:10-zgnr_borondoped+HT_PCD_TotalDOS_PlainBands_SpinDensities}(a), introducing the alternating B$_7$-C$_6$ chain in a 10-ZGNR+H system leads to a dramatically different electronic structure in comparison to the previous cases.
There are three qualitative types of states that contribute to the PCD: localized boron states, delocalized carbon bulk states and localized carbon edge states. The presence of the boron clusters allows both the carbon bulk states and the edge states to hybridize with the cluster induced states. This results in the rather complex and very delocalized PCD shown in Fig.~\ref{fig:10-zgnr_borondoped+HT_PCD_TotalDOS_PlainBands_SpinDensities}(a)

Due to the interplay of the three different types of states the change of the electronic structure in the DOS and band plots in Fig.~\ref{fig:10-zgnr_borondoped+HT_PCD_TotalDOS_PlainBands_SpinDensities} is relatively complicated. But, similar to the previous cases, the DOS near $E_F$ increases, and there are more bands near $E_F$ than in the undoped case due to the appearance of the boron related bands.
In the AFM state the boron doping increases the number of bands at $E_F$ from 0 to 4, and in the FM state the increase is from 2 to 4.
The $\sigma$ states near -1 eV in the DOS of the doped system in Fig.~\ref{fig:10-zgnr_borondoped+HT_PCD_TotalDOS_PlainBands_SpinDensities}(c) are most dispersed in the present system, and they come very close to $E_F$. Further doping of the system could eventually create holes in these kinds of $\sigma$ bands. It is interesting to note that the superconductivity in MgB$_2$ and boron doped diamond is related to holes in $\sigma$ bands \cite{Kortus2001,Boeri2004}. One might speculate whether boron doped ZGRNs could allow for a similar type of superconductivity.

To conclude this section, we found that B$_7$ clusters embedded into graphene systems are structurally stable, and they strongly modify the electronic structure near $E_F$. The boron clusters locally metallize the system. This is achieved both by the reduction of $E_F$, and by the introduction of boron states near $E_F$. A linear chain of boron clusters forms a 1D metallic ``wire'' inside the graphene matrix. In a ZGNR the cluster-related states hybridize with the edge and bulk
states.

\subsection{Energetics and magnetic properties}
\label{sec:The magnetic properties of the boron clusters}
\begin{table}[h!]
\begin{footnotesize}
\caption{The energies, magnetic states and moments of all systems considered in this study. ``State'' denotes different systems and magnetic orderings: UC (unit cell), SC (supercell), H (hydrogen terminated), BD (boron doped), NM (non-magnetic), SP (spin polarized), FM (ferromagnetic), AFM (antiferromagnetic). $N_\mathrm{at}$ is the number of atoms per unit cell. $\mu$ is the total magnetic moment per unit cell. $E_\mathrm{tot}$ is the total energy per unit cell. The magnetic ground state is indicated in boldface.  $\Delta E_\mathrm{mag}$ is defined in Eq.~\ref{eqn:E_mag} and the degree of doping $\Delta E_\mathrm{F}$ is defined in Eq. \ref{eqn:E_Fermi}.}
\label{tab:mag_energies}
\begin{tabular}{llccccc} \hline
System & State &  $N_\mathrm{at}$ &  $E_\mathrm{tot}$ &  $\mu$     & $\Delta E_\mathrm{mag}$ & $\Delta E_\mathrm{F}$ \\
       &       &           &    (eV)    & ($\mu_{B}$)&   (meV/at) & (eV) \\ \hline
Graphene  & UC, NM  & 2  &  -18.479   & - & -  &  - \\ \cline{2-6}
          & SC, NM & 72  &  -664.367 & - &  -  & -\\ \cline{2-6}
          & BD, NM & 73  & -647.617  & -& 9  & 0.38 \\
          & \textbf{BD, SP} & 73  & -647.626  & 0.29& 0  & 0.36 \\ \cline{2-6}
          & iso.-SC, NM & 144  & -1329.832   & -& - & - \\ \cline{2-6}
          & iso.-BD, NM & 145  & -1312.124   & -& 6  & 0.18 \\
          & \textbf{iso.-BD, SP} & 145  & -1312.130 & 1.00& 0  & 0.20 \\  \hline
9-AGNR    & NM & 72  & -648.249 & -&  - & - \\ \cline{2-6}
          & BD, NM & 73 & -631.318 & - & 7  & 0.28 \\
          & \textbf{BD, SP} & 73 & -631.325 & 0.30 & 0  & 0.30 \\ \cline{2-6}
          & H, NM & 80  & -691.588 & - & -  & - \\ \cline{2-6}
          & BD+H, SP & 81 & -674.679 & 0.30 & 6 & 0.29 \\
          & \textbf{BD+H, NM} & 81 & -674.685 & -& 0  & 0.26 \\ \hline
10-ZGNR   & NM & 60  & -535.044 & -& 270 & - \\
          & FM & 60  &-536.629  & 7.68 & 5  & - \\
          & \textbf{AFM} & 60 & -536.661 & 0.00& 0 & - \\\cline{2-6}
          & BD, NM  & 61  & -518.638 & -& 227  & 0.21 \\
          & BD, FM  & 61  &-519.998 & 6.99& 1  & 0.33 \\
          & \textbf{BD, AFM}  & 61 &-520.002 & 0.14& 0 & 0.25 \\ \cline{2-6}
          & H, NM & 66  & -573.242 & -& 27  & - \\
          & H, FM & 66  & -573.368 & 1.37& 6 & - \\
          & \textbf{H, AFM} & 66  & -573.406 & 0.00& 0 & - \\ \cline{2-6}
          & BD+H, NM & 67  & -556.880 & -& 20  & 0.25 \\
          & BD+H, AFM & 67  & -556.995 & 0.22& 1 & 0.37 \\
          & \textbf{BD+H, FM} & 67  & -556.998 & 1.30& 0 & 0.33 \\
\hline
\end{tabular}
\end{footnotesize}
\end{table}

In this section we will investigate the energetics  and magnetic properties of the boron doped graphene systems. The results comprising the energies and the magnetic moments for nonmagnetic (NM) and magnetic (SP, FM, AFM) states are tabulated in Table \ref{tab:mag_energies}. The energies of different magnetic states are characterized by the difference in their total energy as compared to the total energy of the magnetic ground state ($E_\mathrm{tot}^\mathrm{GS}$):
\begin{eqnarray}
\Delta E_\mathrm{mag} = (E_\mathrm{tot} - E_\mathrm{tot}^\mathrm{GS})/N_\mathrm{MA},
\label{eqn:E_mag}
\end{eqnarray}
where $N_\mathrm{MA}$ is the number of magnetic ``centers'' per unit cell. For the majority of systems under consideration  $N_\mathrm{MA}=1$, except for the ZGNR systems, where this value is equal  to the number of carbon edge atoms per unit cell ($N_\mathrm{MA}=6$). When analyzing the spin densities in Figs.~\ref{fig:graphene_borondoped-isolated_PCD_TotalDOS_PlainBands_SpinDensities}(h)-\ref{fig:10-zgnr_borondoped+HT_PCD_TotalDOS_PlainBands_SpinDensities}(h) it becomes quite obvious that the magnetic moments are never concentrated on a single atom. So $N_\mathrm{MA}$ represents the number of similar groups of atoms per unit cell, which collectively carry a certain total magnetic moment.

Undoped graphene and AGNRs are nonmagnetic. However, when graphene is doped with an isolated boron cluster (graphene+isolated-BD) the  system has a total magnetic moment of 1 $\mu_B$ per cluster due to the odd number of electrons per boron cluster (21 in B$_7$ as compared to 24 in C$_6$). Thus a boron cluster carries an unpaired electron that has an magnetic moment of exactly 1 $\mu_B$.
The corresponding spin density shown in Fig.~\ref{fig:graphene_borondoped-isolated_PCD_TotalDOS_PlainBands_SpinDensities}(h) is primarily concentrated on the outer B atoms, and the latter does not interact with the cluster replica in the adjacent unit cells.
By bringing the boron clusters closer to each other within the graphene-BD linear chain system, the magnetic moment is quenched down to 0.3  $\mu_B$/cluster, due to the onset of magnetic exchange interactions between the clusters. The reduction of the magnetic moment is also discernible in the spin density in Fig.~\ref{fig:graphene_borondoped_PCD_TotalDOS_PlainBands_SpinDensities}(h).
The magnetic properties do not change very significantly from the graphene-BD linear chain system to the 9-AGNR+BD(+H) systems. The magnetic moment is still 0.3 $\mu_B$/cluster, and the spin densities of the two systems are relatively similar. This again indicates the the change of the electronic and magnetic structure induced by the presence of the cluster(s) is of a rather local nature.
However there is one thing to be pointed out: the ground state of the 9-AGNR+BD system is magnetic and the ground state of the 9-AGNR+BD+H system is NM. We speculate that the hydrogen passivation removes dangling bonds at the edges, and it also minimizes the exchange interactions that let the system prefer to settle down in a NM ground state.
However, the energy differences between the non-magnetic and the magnetic states ($\Delta E_\mathrm{mag}$) in the graphene and AGNR systems are very small, i.e. between 6-9 meV/atom (Table \ref{tab:mag_energies}).
Such small values indicate that the magnetism in these systems must be very weak.

ZGNRs are believed to carry magnetic moments along the edges. However the magnetic interaction between the two edges of the ribbon is very weak, thus leading to a relatively unstable magnetic system \cite{Kunstmann2011}.
As shown above, the boron cluster usually carries a magnetic moment.
So one might speculate that the magnetic interaction between the edges in ZGNRs could be enhanced by the presence of the boron clusters. The more as both the cluster magnetic moments and the ones on the edges are generated by $\pi$-electrons, and therefore the two systems should be able to interact.

The answer to this question is given in Fig.~\ref{fig:10-zgnr_borondoped+HT_PCD_TotalDOS_PlainBands_SpinDensities}(h). It shows that the edge states are weakened by the presence of the boron cluster. The boron cluster itself does no longer  carry a net magnetic moment. Thus in ZGNR systems the boron cluster locally supresses the magnetic moments, and it effectively weakens the magnetic interactions.
Note that the weakening of the edge magnetism  happens both in the FM and in the AFM state.
Furthermore, for the different FM states that are listed in Tab.~\ref{tab:mag_energies} the total magnetic moment is always smaller in the doped case than in the undoped case.
In the 10-ZGNR system the boron cluster is located at a very asymmetric position in order to break the symmetry between the edges. As result of the asymmetric supression of magnetic moments the positive and negative moments of the AFM states on opposite sides of the ribbon are no longer equal in magnitude, and there is a residual total magnetic moment per cell of 0.1 $\mu_B$/cell and 0.2 $\mu_B$/cell for the unpassivated and H passivated cases, respectively (as opposed to 0 $\mu_B$/cell for the undoped systems).
In the undoped systems the AFM and FM states are energetically separated by only $\Delta E_\mathrm{mag}=5-6$ meV/atom. In the doped systems the two states are isoenergetic with $\Delta E_\mathrm{mag}=1$ meV/atom (smaller than the DFT energy resolution). This again is a strong indication that magnetic interactions are suppressed by the presence of the boron cluster.
Note that a similar suppression of the edge magnetic moments with increasing Ti concentration near the edge of ZGNRs  is reported in Ref. \cite{Guodong2012}.

We can summarize this section by saying that the magnetism in boron doped graphene systems is generally very weak. Furthermore the edge magnetism in ZGNRs in weakened by the presence of boron clusters.

\section{Summary and Conclusion}
\label{sec:Conclusion}

We studied the doping of graphene and various graphene nanoribbons by embedded B$_7$ clusters. The embedding seems to be stable, and the boron clusters tend to strongly modify the electronic structure near $E_F$, which leads to a very local metallization of the graphene substrates. This metalliation is achieved by both the reduction of $E_F$, and by the introduction of boron states near $E_F$.

We also showed that a linear chain of boron clusters forms an effective 1D metallic ``wire'' inside a semiconducting substrate made of a suitable graphene nanoribbon, where the metallicity rapidly decays away from the wire. Given the fact that a semiconducting graphene substrate does not require any additional doping, any controlled layout of linear chains of boron clusters might actually be the technological basis of a layout of integrated circuits, which is at least one or two  length scales below state--of--the--art microelectronics.  Note that the corresponding cluster implantation technique can already be applied to graphene.

Furthermore the presence of the boron clusters also introduces $\sigma$ states in the occupied part of the band structure within 1 eV below the Fermi level. We wonder whether this particular feature could give rise to electron-phonon mediated superconductivity, similar to MgB$_2$ or boron doped diamond.

Finally, the magnetism in boron doped graphene systems is generally very weak, and the edge magnetism in ZGNRs is further weakened by the presence of boron clusters. Therefore we do not foresee any application of embedded boron clusters in the field of graphene based spintronics, supposing that such a technology will be feasible at all in carbon nanosystems.

We hope that our results might also lead to an incresing interest in boron based nanomaterials, whose polymorphism is at least as rich as the polymorphism of carbon nanomaterials \cite{Quandt2005,Bezugly2011,Boustani2011a}. Our present results indicate that interesting functionalities might actually be gained from the combination of carbon and boron nanosystems, and this might be a promising new field to be explored in the future, besides pure graphene.

\begin{acknowledgments}
J.K. acknowledges financial support from the DFG (project KU 2347/2-2).
The computations were partly performed at the computation facility of \c{C}ankaya University, at the ULAKB\.IM High Performance Computing Center at the Scientific and Technological Research Council of Turkey and at The Center for Information Services and High Performance Computing (ZIH) of  the TU Dresden. A. Q. acknowledges financial support by the National Research Foundation (NRF) and the Department of Science and Technology (DST) of South Africa, as well as by the Gauteng node of the National Institute of Theoretical Physics (NITheP). We further acknowledge the German Excellence Initiative via the Cluster of Excellence 1056 “Center for Advancing Electronics Dresden” (cfAED).
\end{acknowledgments}

\bibliography{paper}

\end{document}